\documentclass[reprint,amsmath,amssymb,aps,prl,superscriptaddress]{revtex4-2}

\usepackage{tikz-cd}

\DeclareMathOperator{\Gr}{Gr}
\DeclareMathOperator{\Tr}{Tr}

\begin{document}

\title{Cluster Algebras, Cube Roots, and Energy Correlators in $\mathcal{N}=4$ SYM}

\author{Dani Kaufman}
\affiliation{Max Planck Institute for Mathematics in the Sciences, 04103 Leipzig, Germany}

\author{Marcus Spradlin}
\affiliation{Department of Physics, Brown University, Providence, RI 02912, USA}
\affiliation{Brown Center for Theoretical Physics and Innovation, Brown University, Providence, RI 02912, USA}

\author{Anastasia Volovich}
\affiliation{Department of Physics, Brown University, Providence, RI 02912, USA}

\author{Kai Yan}
\affiliation{School of Physics and Astronomy, Shanghai Jiao Tong University, Shanghai 200240, China}
\affiliation{Key Laboratory for Particle Astrophysics and Cosmology (MOE), Shanghai 200240, China}

\begin{abstract}
We provide a cluster algebraic construction of the cube root symbol letters that appear at leading order in the near-collinear expansion of the four-point energy correlator in $\mathcal{N}=4$ SYM theory.
\end{abstract}

\maketitle

\section{Introduction}

Scattering amplitudes in quantum field theory can have very intricate analytic structure. In ``simple'' theories like $\mathcal{N}=4$ super-Yang-Mills (SYM) it is observed and expected that this structure should be dictated by geometric and combinatoric considerations: for example, in some of the simplest cases, by cluster algebras~\cite{Golden:2013xva} (see the book~\cite{FWZbook} for an introduction). While the singularity structure of general amplitudes can seemingly be arbitrarily complicated (see~\cite{Chestnov:2026mpo}), it has been noted in recent years that cluster algebras also underlie the singularity structure of some amplitudes and individual Feynman integrals well beyond SYM theory~\cite{Chicherin:2020umh,Bossinger:2022eiy,Aliaj:2024zgp,Pokraka:2025ali,Bossinger:2025rhf,Aliaj:2026iny}.

It is interesting to ask whether similar structures might underlie the singularities of other observables, such as energy correlators, which have received increasing attention in recent years (see~\cite{Moult:2025nhu} for a review). One especially interesting feature of the four-point energy correlator E${}^4$C in SYM theory, which was recently computed at leading order in the multi-collinear limit~\cite{Chicherin:2024ifn}, is the appearance of singularities that are degree-3 algebraic functions of the relevant kinematic variables. No cube roots have been found (as of yet) in the forefront of amplitude calculations in SYM theory, although they do appear in individual Feynman diagrams; see for example~\cite{Bourjaily:2022vti}.

While cube roots are new and exciting, and the primary motivation for this paper, square roots of kinematic variables are ubiquitous in quantum field theory. While cluster variables are always rational functions of those in the initial cluster, there are several closely related ways of associating degree-2 algebraic functions to (infinite) cluster algebras; see for example~\cite{Arkani-Hamed:2019rds,Henke:2019hve,Herderschee:2021dez}, and in particular~\cite{Drummond:2019cxm} which we review in the next section.

In this paper we first note that the symbol letters~\cite{Goncharov:2010jf} of the leading-order near-collinear limit of the three-point energy correlator E${}^3$C in SYM theory are dictated by an $A_2$ cluster algebra. Then we identify a particular cluster algebra seed whose infinite mutation sequences precisely encode the singularities in the cubic algebraic letters of the near-collinear limit of the E${}^4$C. We end with a discussion of the mathematics underlying our results and suggest a generalization to arbitrary degree.

\section{Cluster Algebras and the E${}^3$C}

Let us begin with a simple observation about the three-point energy correlator in SYM theory and the $A_2$ cluster algebra. This correlator was computed to leading order in the collinear limit in~\cite{Chen:2019bpb} and the result involves polylogarithm functions of weight up to 2 with the five-letter symbol alphabet
\begin{align}
\label{eq:eeecalphabet}
    \{z, \overline{z}, 1-z, 1-\overline{z}, 1-|z|^2\}\,,
\end{align}
where $z$ is related to the complex parameters $z_i$ that specify the locations of the three detector operators on the celestial sphere by
\begin{align}
    z \overline{z} = \frac{|z_{23}|^2}{|z_{12}|^2}\,, \qquad
    (1 - z)(1 - \overline{z}) = \frac{|z_{13}|^2}{|z_{12}|^2},
\end{align}
with $z_{ij} = z_i - z_j$. Now if we reparameterize by introducing two variables
\begin{align}
    x_1 = \frac{z \overline{z}}{z - z \overline{z}}\,, \quad
    x_2 = \frac{1 - z}{z - z \overline{z}}
\end{align}
then~(\ref{eq:eeecalphabet}) is easily seen to be equivalent (that means equal, after an invertible multiplicative
transformation) to
\begin{align}
    \{ x_1, x_2, \frac{1+x_2}{x_1}, \frac{1+x_1}{x_2}, \frac{1+x_1+x_2}{x_1 x_2}\}
\end{align}
which are the cluster variables of the $A_2$ cluster algebra.

\section{Square Root Letters from Cluster Algebras}

The discussion here is essentially equivalent to that of Section~3 of~\cite{Drummond:2019cxm}, but we have adapted the notation and presentation in order to align with our following section. Consider the quiver
\begin{equation}
Q_2 := \vcenter{\hbox{
\begin{tikzcd}[row sep=0em]
  & b \arrow[dd, Rightarrow] 
  & \\
c \arrow[ur] 
  & 
  & d \arrow[ul] \\
  & a \arrow[ul] \arrow[ur]
\end{tikzcd}
}}.
\end{equation}
If we mutate on node $b$ and then flip the quiver over the horizontal axis, it comes back to itself with the variables $a$ and $b$ replaced by
\begin{align}
\label{eq:mu2}
    \mu_{b} : (a, b) \mapsto \left( \frac{a^2 + c d}{b}, a\right).
\end{align}
The inverse of this is
\begin{align}
    \mu_{b}^{-1} : (a,b) \mapsto \left(b, \frac{b^2 + c d}{a}\right).
\end{align}
Note that the quantity
\begin{align}
    F = \frac{a^2 + b^2 + c d}{ab}
\end{align}
is invariant under this operation: $\mu_{b}(F) = F$; this is an example of a mutation invariant function in the sense of~\cite{KaufmanMutationInvariantFunctions2024}. Specifying values $(a^{(0)}, b^{(0)})$ in the initial quiver, as well as fixed values for $c$ and $d$ determines the infinite sequences
\begin{align}
    (a^{(n)}, b^{(n)}) := \mu_{b}^n (a^{(0)}, b^{(0)})\,, \qquad n \in \mathbb{Z}\,.
\end{align}
The mutation rule~(\ref{eq:mu2}) gives a recursion that can be solved exactly, with $b^{(n)} = a^{(n-1)}$ and
\begin{align}
\label{eq:zn}
    a^{(n)} = \omega_+ \lambda^n + \omega_- \lambda^{-n}
\end{align}
in terms of
\begin{align}
\label{eq:quadratic}
    \lambda = \frac{1}{2} (F + \sqrt{F^2 - 4})\,, ~~
    \omega_\pm = \frac{a}{2} \left[ 1 \pm  \frac{F - \frac{2 b}{a}}{\sqrt{F^2 - 4}}\right].
\end{align}
According to the proposal of~\cite{Drummond:2019cxm}, we identify the combination $\omega_-/\omega_+$ as an algebraic letter associated to this infinite mutation path. Two additional such letters can be obtained by considering the infinite sequence mutating first on $a$ and then on $b$. This would lead to the same expressions~(\ref{eq:quadratic}) but with $a$ and $b$ exchanged; let us denote the corresponding coefficients
\begin{align}
    \widetilde{\omega}_\pm = \frac{b}{2} \left[ 1 \pm \frac{F - \frac{2 a}{b}}{\sqrt{F^2-4}}\right].
\end{align}
Including the initial variables $a, b$, altogether the symbol letters associated to this construction are therefore
\begin{align}
    \label{eq:quadraticalphabet1}
    \Big\{ a, b, \frac{\omega_-}{\omega_+}, \frac{\widetilde{\omega}_-}{\widetilde{\omega}_+} \Big\}.
\end{align}

Quadratic algebraic letters of this type are well-known in physics and appear for example in the one-loop four-mass box integral. The symbol alphabet of this integral can be expressed (see for example~\cite{Spradlin:2011wp}) as
\begin{align}
\label{eq:quadraticalphabet2}
    \Big\{w, \overline{w}, 1 - w, 1 - \overline{w}\Big\}
\end{align}
where
\begin{align}
\label{eq:uvdef}
    w \overline{w} = u\,, \qquad (1 - w)(1 - \overline{w}) = v
\end{align}
in terms of two conformal cross-ratios $u, v$. Under the identification
\begin{align}
    u = - \frac{a^2}{c d}\,, \qquad v = - \frac{b^2}{c d}\,,
\end{align}
one can check that the identities
\begin{align}
    w^2 = \frac{a^2}{cd} \left(\frac{\widetilde{\omega}_-}{\widetilde{\omega}_+}\right)^{\pm 1}\,, \qquad
    (1 - w)^2 = \frac{b^2}{c d} \left( \frac{\omega_+}{\omega_-}\right)^{\pm 1}
\end{align}
and their conjugates hold, where the power depends on the branch choice for solving~(\ref{eq:uvdef}). In either case, the two alphabets~(\ref{eq:quadraticalphabet1}) and~(\ref{eq:quadraticalphabet2}) are related by an invertible multiplicative transformation, and hence equivalent. In this sense the letters associated to the quiver $Q_2$ are precisely those of the four-mass box function. In~\cite{Drummond:2019cxm} an analysis of infinite mutation sequences associated to all quivers of type $Q_2$ inside the $\Gr(4,8)$ cluster algebra revealed that they encode the complete (to date) set of 18 independent quadratic algebraic symbol letters that are known~\cite{He:2019jee,Li:2021bwg} to appear in eight-particle amplitudes in SYM theory.

\section{Cube Root Letters from Cluster Algebras}

Consider the quiver
\begin{equation}
Q_3 := \vcenter{\hbox{
\begin{tikzcd}[row sep=0em]
  & b \arrow[dd, Rightarrow] 
  & c \arrow[l] \arrow[dr] 
  & \\
e \arrow[ur] 
  & 
  & 
  & f \arrow[dl] \\
  & a \arrow[r] \arrow[ul] 
  & d \arrow[uu, Rightarrow] 
\end{tikzcd}
}}.
\end{equation}
Applying the pair of mutations $\mu_{bd} := \mu_b \mu_d = \mu_d \mu_b$ gives
\begin{equation}
\mu_{bd}(Q_3) =
\vcenter{\hbox{
\begin{tikzcd}[row sep=0em]
  & \frac{a^2 + e c}{b} \arrow[r] \arrow[dl]
  & c \arrow[dd, Rightarrow]
  & \\
e \arrow[dr]
  &
  &
  & f \arrow[ul] \\
  & a \arrow[uu, Rightarrow]
  & \frac{a f + c^2}{d} \arrow[l] \arrow[ur]
\end{tikzcd}
}}
\end{equation}
which, after flipping over the horizontal axis, is seen to have exactly the same form as $Q_3$ but with the variables transformed according to
\begin{align}
  \mu_{bd} :  (a,b,c,d) \mapsto \left( \frac{a^2 + e c}{b}, a, \frac{a f + c^2}{d}, c \right).
\end{align}
The inverse of this is
\begin{align}
    \mu_{bd}^{-1} : (a,b,c,d) \mapsto \left( b, \frac{b^2 + d e}{a}, d, \frac{bf +d^2}{c}\right).
\end{align}
There are precisely two independent quantities, given by
\begin{equation}
\begin{aligned}
F_1 &= \frac{c^2 d e + b^2 c^2 + a^2 c d + a b^2 f  + a b d^2 }{a b c d}\,, \\
F_2 &= \frac{b^2 c d + a b c^2 + a^2 bf + c d^2 e + a^2 d^2}{a b c d}
\end{aligned}
\end{equation}
that are invariant under this operation:
\begin{equation}
    \mu_{bd}(F_i) = F_i\,.
\end{equation}

Specifying values $(a^{(0)}, b^{(0)}, c^{(0)}, d^{(0)})$ in the initial quiver, as well as fixed values for $e$ and $f$ determines the infinite sequences
\begin{align}
    (a^{(n)}, b^{(n)}, c^{(n)}, d^{(n)}) := \mu_{bd}^n (a^{(0)}, b^{(0)}, c^{(0)}, d^{(0)})\,.
\end{align}
Our aim is to analyze the asymptotic growth of these variables as $|n| \to \infty$. Since $b^{(n)} =  a^{(n-1)}$ and $d^{(n)} = c^{(n-1)}$ it is sufficient to focus on the $a$ and $c$ variables. To that end we first note that these satisfy the coupled quadratic recursion
\begin{align*} 
    a^{(n)}c^{(n)}F_1 &= a^{(n+1)}c^{(n)}+a^{(n-1)}c^{(n+1)}+a^{(n)}c^{(n-1)}\,,\\
    a^{(n)}c^{(n)}F_2 &= a^{(n)} c^{(n+1)}+a^{(n+1)}c^{(n-1)}+a^{(n-1)}c^{(n)}\,.
\end{align*}
It is straightforward to check that these can be solved explicitly by making the ansatz
\begin{equation}
\begin{aligned}
    F_1 &= \Tr[M] = \lambda_1 +\lambda_1^{-1}\lambda_2+\lambda_2^{-1}\,, \\
    F_2 &= \Tr[M^{-1}] = \lambda_2 +\lambda_2^{-1}\lambda_1+\lambda_1^{-1}\,,
\end{aligned}
\end{equation}
where
\begin{align}
\label{eq:Mdef}
    M = \begin{bmatrix}\lambda_1 &0 &0 \\ 0& \lambda_1^{-1}\lambda_2 & 0 \\ 0& 0& \lambda_2^{-1}\end{bmatrix},
\end{align}
together with
\begin{equation}
\begin{aligned}
    a^{(n)} &= \omega_1\, \lambda_1^{n} +\omega_2\, \lambda_1^{-n}\lambda_2^n+\omega_3\, \lambda_2^{-n}\,,\\
    c^{(n)} &= \omega_4\, \lambda_2^{n} +\omega_5\, \lambda_2^{-n}\lambda_1^n+\omega_6\, \lambda_1^{-n}\,,
\end{aligned}
\end{equation}
with the six $\omega_i$ determined in terms of triples of initial conditions, say $[a_{-1}, a_0, a_{1}]$ and $[c_{-1}, c_0, c_{1}]$, via
\begin{equation}
\begin{aligned}
\label{eq:omegas}
    \begin{bmatrix}
        \omega_1\\\omega_2\\ \omega_3 
    \end{bmatrix} 
    = \begin{bmatrix}
        \lambda_1 &\lambda_1^{-1}\lambda_2& \lambda_2^{-1} \\
        1 & 1 & 1 \\
     \lambda_1^{-1} & \lambda_2^{-1}\lambda_1& \lambda_2
    \end{bmatrix}^{-1}\begin{bmatrix}
        a_{1}\\a_0\\ a_{-1} 
    \end{bmatrix}, \\
     \begin{bmatrix}
        \omega_4\\\omega_5\\ \omega_6 
    \end{bmatrix} 
    = \begin{bmatrix}
        \lambda_2 &\lambda_2^{-1}\lambda_1& \lambda_1^{-1} \\
        1 & 1 & 1 \\
     \lambda_2^{-1} & \lambda_1^{-1}\lambda_2& \lambda_1
    \end{bmatrix}^{-1}\begin{bmatrix}
        c_{1}\\c_0\\ c_{-1} 
    \end{bmatrix}.
\end{aligned}
\end{equation}
When the initial cluster variables are evaluated at positive real numbers, the characteristic polynomial
\begin{align}
\label{eq:monic_poly}
    p_M(x) = x^3 - F_1 x^2 + F_2 x - 1
\end{align}
has distinct positive real roots (see discussion below). We choose to sort these roots so that
\begin{align}
\label{eq:sorted}
    \lambda_1>\lambda_1^{-1}\lambda_2>\lambda_2^{-1}\,,
\end{align}
in which case the asymptotic behavior of the cluster variables is
\begin{equation}
\begin{aligned}
    a^{(n)} \sim \begin{cases} \omega_1 \lambda_1^n & n \to + \infty \\ \omega_3 \lambda_2^{-n} & n \to -\infty \end{cases}, \\
    c^{(n)} \sim \begin{cases} \omega_4 \lambda_2^n & n \to + \infty \\ \omega_6 \lambda_1^{-n} & n \to -\infty \\ \end{cases}.
\end{aligned}
\end{equation}
This would suggest that we consider the asymptotic coefficients $\omega_1$, $\omega_3$, $\omega_4$ and $\omega_6$ as algebraic letters associated to the infinite mutation path. However, we shall see in the next section that the natural domain for the physics problem under consideration is not the one where all cluster variables are positive, or even in general real; instead the $\lambda_i$ may be complex and we will consider all six $\omega_i$ on equal footing and view them collectively as cubic algebraic letters associated to the mutation class of $Q_3$.

We shall need explicit expressions for the $\omega_i$ which encode the symbol letters we are after. With the benefit of hindsight, let us choose to parameterize the variables $(a,b,c,d)$ in the initial quiver in terms of four variables $(u_1,u_2,u_3,v)$ as
\begin{equation}
\begin{aligned}
a &=  E^{\frac23}F^{\frac13}\, v(v^2-u_1u_2u_3)\,, \\
b &=  E^{\frac23}F^{\frac13}\, u_1^{\frac23}u_2^{\frac23}u_3^{\frac23}(v^2-u_1u_2u_3)\,, \\
c &=  E^{\frac13}F^{\frac23}\, v(v^2-u_1u_2u_3)\,,  \\
d &=  E^{\frac13}F^{\frac23}\, u_1^{\frac13}u_2^{\frac13}u_3^{\frac13}(v^2-u_1u_2u_3)\,, 
\end{aligned}
\end{equation}
where $E$ and $F$ stand for
\begin{equation}
\begin{aligned}
E&= \frac{e}{(u_1 u_2 -v )  (u_2 u_3-v )  (u_1 u_3-v ) }\,, \\
F&= \frac{f}{(u_1 -v)  (u_2 -v)  (u_3-v) } \,.
\end{aligned}
\end{equation}
In this parameterization the invariants evaluate to
\begin{align}
    F_1 = \frac{u_1 u_2 + u_1 u_3 + u_2 u_3}{u_1^\frac23 u_2^\frac23 u_3^\frac23}\,, \quad
    F_2 = \frac{u_1 + u_2 + u_3}{u_1^\frac13 u_2^\frac13 u_3^\frac13}
\end{align}
and we can choose
\begin{align}
\label{eq:lambdau}
\lambda_1 = u_1^{-\frac23} u_2^{\frac13} u_3^\frac13\,, \quad
\lambda_2 = u_1^{-\frac13} u_2^{-\frac13} u_3^\frac23\,.
\end{align}
The $\omega_i$ can then be calculated from~(\ref{eq:omegas}), and we find
\begin{equation}
\begin{aligned}
w_1 &=
 E^{2/3} F^{1/3}\,
\frac{
u_2 u_3 (u_1-v)(u_1u_2{-}v)(u_1u_3{-}v)
}{
(u_1-u_2)(u_3-u_1)
}\,, \\
w_2 &=
E^{2/3} F^{1/3}\,
\frac{
u_1 u_3 (u_2-v)(u_1u_2{-}v)(u_2u_3{-}v)
}{
(u_1-u_2)(u_2-u_3)
}\,, \\
w_3 &=
 E^{2/3} F^{1/3}\,
\frac{
u_1 u_2 (u_3-v)(u_1u_3{-}v)(u_2u_3{-}v)
}{
(u_3-u_1)(u_2-u_3)
}\,, \\
w_4 &=
E^{1/3} F^{2/3}\,
\frac{
u_3^2 (u_1-v)(u_2-v)(u_1u_2{-}v)
}{
(u_3-u_1)(u_2-u_3)
}\,, \\
w_5 &=
E^{1/3} F^{2/3}\,
\frac{
u_2^2 (u_1-v)(u_3-v)(u_1u_3{-}v)
}{
(u_1-u_2)(u_2-u_3)
}\,, \\
w_6 &=
 E^{1/3} F^{2/3}\,
\frac{
u_1^2 (u_2-v)(u_3-v)(u_2u_3{-}v)
}{
(u_1-u_2)(u_3-u_1)
}\,.
\end{aligned}
\end{equation}
Altogether this construction generates symbol letters containing 12 factors involving cubic algebraic functions:
\begin{multline}
\label{eq:cubicletters1}
    \{u_1,u_2,u_3, u_2 - u_1, u_3 - u_1, u_3 - u_2, u_1-v, \\
    u_2-v, u_3-v, u_1 u_2-v, u_1u_3-v, u_2u_3-v \}\,.
\end{multline}

\section{Cube Roots in the E${}^4$C}

The collinear limit of the four-point energy correlator in $\mathcal{N}=4$ super Yang-Mills~\cite{Chicherin:2024ifn} theory depends on two complex parameters $Z, W$ related by
\begin{equation}
\begin{aligned}
    Z \overline{Z} &= \frac{|z_{12}|^2}{|z_{13}|^2}\,, &  (1-Z)(1-\overline{Z}) &= \frac{|z_{23}|^2}{|z_{13}|^2}\,,\\
    W \overline{W} &= \frac{|z_{24}|^2}{|z_{34}|^2}\,, &  (1-W)(1-\overline{W}) &= \frac{|z_{23}|^2}{|z_{34}|^2}
\end{aligned}
\end{equation}
to the locations $z_i$ of the four detector operators. At leading order the correlator involves polylogarithm functions of weight up to 3 with a symbol alphabet consisting of four different types of letters. Some letters, such as $1 - |Z|^2$ and $|Z|^2 - |W|^2$, are rational functions of the $|z_{ij}|^2$. Others, such as $Z$ and $1 - W$, are quadratic algebraic functions. Still others, such as
\begin{align}
    \frac{a + W}{b + W} \frac{b + \overline{W}}{a + \overline{W}}\,,
\end{align}
where $a,b,c$ are the roots of the cubic polynomial
\begin{multline}
   p_3(x)= x^3 + ( W + \overline{W} + Z + \overline{Z}-1)x^2 + |W|^2|Z|^2 \times \\
\big( Z^{-1}+\overline{Z}^{-1} + W^{-1}+\overline{W}^{-1} - 1\big)x
+ |W|^2|Z|^2\,,
\label{eq:cubic}
\end{multline}
are sextic algebraic functions; we defer consideration of these more complicated letters for future work. (We caution the reader that the roots $a, b, c$ of the cubic polynomial $p_3$ in this section have no simple relation to the cluster variables $a, b, c$ labeling the nodes on the quiver $Q_3$ in the previous section.)

Our focus is on the six independent letters that are degree-3 algebraic functions of the $|z_{ij}|^2$:
\begin{align}
\label{eq:cubicletters2}
    \left\{ \frac{a}{b}, \frac{b}{c}, \frac{a + |W|^2}{b + |W|^2}, \frac{b + |W|^2}{c + |W|^2},
    \frac{a + |Z|^2}{b + |Z|^2}, \frac{b + |Z|^2}{c + |Z|^2}
    \right\}.
\end{align}
Note that thanks to the identity
\begin{align}
\frac{a+|Z|^2}{b+ |Z|^2} = \frac{a(b c {-}|W|^2)}{b(a c {-} |W|^2)} 
\end{align}
(and its cyclic partners) the alphabet~(\ref{eq:cubicletters2}) can be traded for the multiplicatively equivalent set
\begin{align}
\label{eq:cubicletters3}
    \left\{ \frac{a}{b}, \frac{b}{c}, \frac{a + |W|^2}{b + |W|^2}, \frac{b + |W|^2}{c + |W|^2},
    \frac{a b {-} |W|^2}{b c {-} |W|^2},
    \frac{b c {-} |W|^2}{a c {-} |W|^2}
    \right\}.
\end{align}
Upon identifying
\begin{align}
    (u_1, u_2, u_3, v) = (-a, -b, -c, |W|^2)
\end{align}
we immediately see that every factor in~(\ref{eq:cubicletters3}) is contained in the list~(\ref{eq:cubicletters1}). The only factors missing---that are absent from the E${}^4$C but generated by the cluster algebra---are those of the form $u_i - u_j$. Symbol letters of this type would indicate logarithmic singularities at the locus when two roots of the cubic~(\ref{eq:cubic}) collide.

\section{Mathematical Discussion}

The cluster algebras considered in the computation of the quadratic and cubic roots shown here are related to Fock-Goncharov moduli spaces of $G$ local systems on surfaces~\cite{FockGoncharovModuliSpacesLocal2006}. Specifically, the seeds $Q_2$ (resp. $Q_3$) come from the spaces of decorated $G=\mathrm{SL}_2$ (resp. $G=\mathrm{SL}_3$) local systems on an annulus. The local system determines a single element of $G$ which is well-defined up to conjugation. The cluster variables can be used to compute an element $M\in G$ in this conjugacy class; its entries are given explicitly as rational functions in the cluster variables. In the cubic case, the functions $F_1,F_2$ encode the two conjugation invariant functions on elements of $\mathrm{SL}_3$ matrices, namely $\Tr(M)$ and $\Tr(M^{-1})$. Fock and Goncharov prove that when the cluster variables are evaluated over the positive real numbers these matrices always have distinct positive eigenvalues, justifying our assumption that $M$ has the form stated in~(\ref{eq:Mdef}).

The mutation sequence we consider realizes the mapping class group action on the annulus, i.e. it acts via a Dehn twist around the simple closed curve. This helps explain why the growth of the cluster variables is determined by the eigenvalues of $M$. 

\subsection{Realizing $Q_3$ inside $\Gr(4,16)$ via mutation invariants}

\begin{figure}
    \centering
    \includegraphics[width=0.95\linewidth]{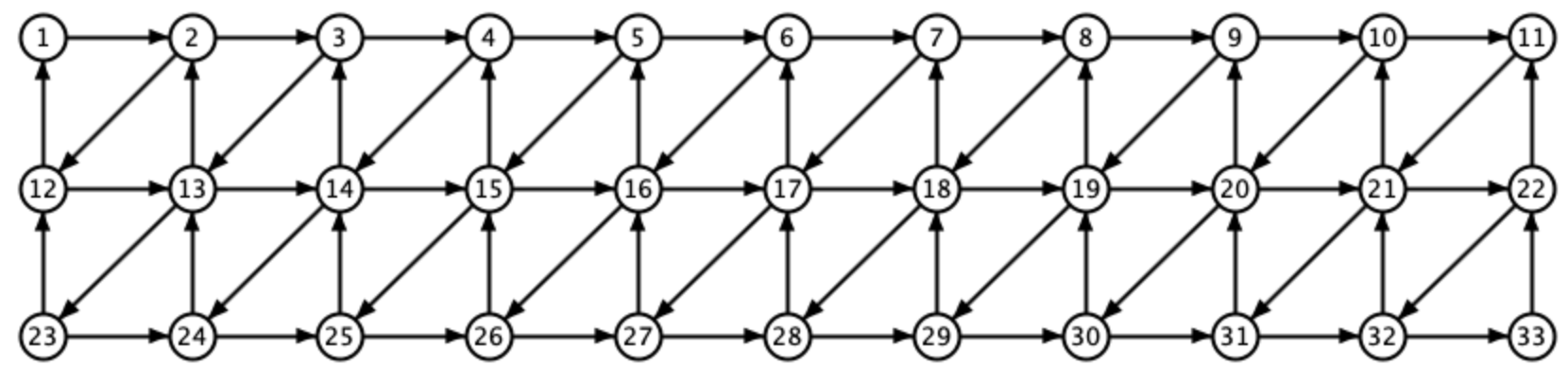}
    \caption{Initial seed for $\Gr(4,16)$.}
    \label{fig:gr416}
\end{figure}

\begin{figure}
    \centering
    \includegraphics[width=0.95\linewidth]{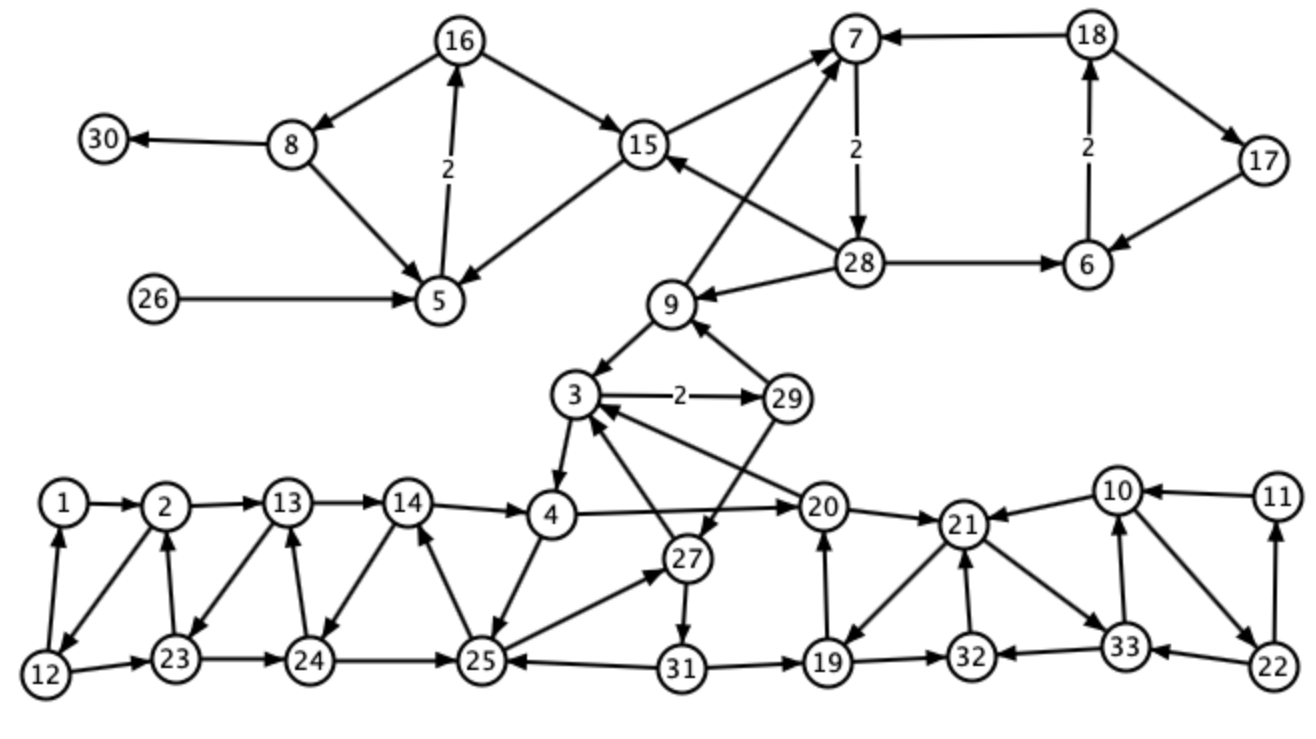}
    \caption{After applying the mutation sequence in the text.}
    \label{fig:gr416mutated}
\end{figure}

Starting with the seed for the Grassmannian $\Gr(4,16)$ shown in Fig.~\ref{fig:gr416} and applying the mutation sequence on the following nodes (read left to right): 33 22 11 32 21 31 20 10 33 22 32 9 21 33 11 10 22 11 23 33 21 13 24 32 4 14 9 20 31 19 8 19 8 20 31 20 29 19 15 25 31 29 8 29 8 28 18 8 29 31 20 17 5 4 17 27 17 7 29 8 9 7 17 27 8 29 9 29 8 17 7 29 8 17 7 6 18 7 9 3 7 9 8 18 8 18 16 28 17 15 29 8 5 27 16 18 9 6 29 28 7 15 18 16 18 16 28 16 28 7 29 28 7 28 15 9 18 7 15 6 15 6 28 7 18 16 15 one obtains Fig.~\ref{fig:gr416mutated}, which contains quiver $Q_3$ on the nodes 6, 7, 15, 17, 18, 28.

We found this mutation sequence by using the \textit{cluster braid symmetries}~\cite{FraserBraidGroupSymmetries2020} of the Grassmannian cluster algebra, or more specifically their categorification, see \cite{Drummond:2026gzt}. The Grassmannians $\Gr(4,4m)$ have 3 braid symmetry generators $\sigma_1,\sigma_2,\sigma_3$. The mutation sequence given above was found by computing a maximal collection of mutation invariant cluster variables under the symmetry $\tau^8:=(\sigma_1\sigma_2\sigma_1)^8=(\sigma_2\sigma_1\sigma_2)^8$. In fact, in the final cluster all variables other than those on nodes 28, 6, 7, 18 are invariant under $\tau^8$ and the mutation sequence $\mu_{bd}$ we considered earlier acts as the symmetry $\tau^8$.

There is an interpretation for functions that are mutation invariant under the element $\tau$ as opposed to $\tau^8$. Recall that the cross-ratios of the Grassmannian cluster algebra parameterize $\mathrm{PGL}_4$ configurations of $n$ points (momentum twistors in physics) $Z_1,\dots ,Z_n \in \mathbb{P}^3$. When $n=16$ we can make a configuration of 4 points and 4 planes by considering the points $Z_1,Z_5,Z_9,Z_{13}$ and the planes through $\{Z_2,Z_3,Z_4\},\dots,\{Z_{14},Z_{15},Z_{16}\}$. (One might think of this as a configuration of 4 twistor points and 4 dual twistor points.) This space is 9-dimensional and the cross ratios ($\mathcal{X}$-coordinates) at nodes 30, 26, 8, 5, and 16 in Fig.~\ref{fig:gr416mutated}, the two invariant monomials in cross ratios $x_{15}^3x_7^2x_{28}^2x_{18}x_6,\, x_{17}^3x_6^2x_{18}^2x_{28}x_7$ along with the two invariants $F_1,F_2$ of the action of $\tau^8$ give a basis for its function field. 

This is exactly a generalization of the quadratic case in $\Gr(4,8)$; there the quiver $Q_2$ is realized by computing a maximal collection of cluster variables under the action $(\sigma_1)^4=(\sigma_3)^4$ and the mutation sequence $\mu_{ab}$ we considered there realizes this symmetry. The set of invariants under $\sigma_1\sigma_3$ is two dimensional and parameterizes configurations of 4 lines in $\mathbb{P}^3$, which is exactly the kinematic space for the four-mass box function.

\subsection{Generalization}

\begin{figure}
    \centering
\begin{tikzcd}[row sep=0em] & a_2 \arrow[dd, Rightarrow] & a_3 \arrow[l] \arrow[r]    & a_6 \arrow[dd, Rightarrow] \arrow[r, no head, dashed] & a_{2n-2} \arrow[rd]             &                \\
f_1 \arrow[ru]  &                            &                            &                                                       &                                 & f_2 \arrow[ld] \\
                & a_1 \arrow[r] \arrow[lu]   & a_4 \arrow[uu, Rightarrow] & a_5 \arrow[l] \arrow[r, no head, dashed]              & a_{2n-3} \arrow[uu, Rightarrow] &               
\end{tikzcd}
    \caption{A candidate cluster $Q_n$.}
    \label{fig:generalization}
\end{figure}
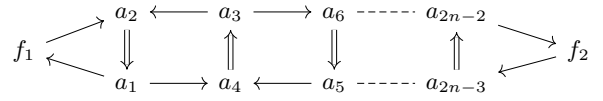

It is now clear how one might generalize the story here. The cluster algebra associated to $\mathrm{SL}_n$ local systems on an annulus contains the cluster algebra shown in Fig.~\ref{fig:generalization} (generally as a subalgebra). Then the cluster symmetry which acts by mutation at the even indexed variables followed by horizontal reflection is a realization of a Dehn twist on the annulus. One could compute its invariants and give a general formula for the cluster variables found along this mutation sequence and then hope that these quantities are related to higher order algebraic roots.

\section{Outlook}

The multi-collinear limit of the four-point energy correlator in SYM theory presents an interesting example of a quantum field theory observable whose singularity structure (specifically, in this case, symbol alphabet) involves algebraic functions of degree 3 in the fundamental kinematic parameters, in this case the positions $z_i$ of the four detector operators. We have identified a particular cluster algebra seed $Q_3$ whose infinite mutation sequences encode all of these letters. It would be very interesting to find a cluster algebra, possibly with a seed containing $Q_3$ as a subquiver, that encodes all symbol letters of the E${}^4$C including those of degree 1, 2, and 6.

Our result opens several avenues for future work. Naturally it would be interesting to systematically map out the algebraic letters associated to other infinite mutation sequences in the algebra seeded by $Q_3$, and to identify other quivers leading to cubic or higher-order roots. Unlike for amplitudes in SYM theory, where the momentum twistor parameterization of the kinematic space indicates a clear role for the $\Gr(4,n)$ cluster algebra~\cite{Golden:2013xva}, it is not clear what the natural cluster algebra for energy correlators ``should'' be.  We have checked that there is a cluster of $\Gr(4,16)$ that contains a subquiver in the shape of $Q_3$; it is possible that it exists also in $\Gr(4,n)$ for $n$ slightly smaller than 16. We expect that clarifying this structure will help identify the natural variables and bootstrap constraints for higher-point energy correlators, much as cluster algebras have done for scattering amplitudes. 

\section{Acknowledgments}

We are grateful to H.~Thomas and H.-C.~Weng for helpful discussions. This work was supported in part by the US Department of Energy under contract DE-SC0010010 Task F (MS, AV), Simons Investigator Award \#376208 (AV), and the National Natural Science Foundation of China under Grant No.~12357077 (KY). DK, MS and AV would like to thank the Erwin Schr\"odinger International Institute for Mathematics and Physics (ESI) at the University of Vienna (Austria) for the opportunity and financial support to participate in the Thematic Programme ``Amplitudes and Algebraic Geometry" in 2026, where this work was initiated. This work was performed in part at the Aspen Center for Physics, which is supported by National Science Foundation grant PHY-2210452.

\bibliographystyle{apsrev4-2}
\bibliography{draft.bib}

\end{document}